\documentclass[sigconf,natbib=true]{acmart}

\AtBeginDocument{%
  }

\acmSubmissionID{123-A56-BU3}

\usepackage{enumitem}         % 调整列表间距
\usepackage{threeparttable}   % 表格注释
\usepackage{siunitx}          % 表格数字对齐 (S列)
\usepackage{tikz}             % 绘图
\usepackage{pgfplots}         % 统计图表
\pgfplotsset{compat=1.17}     % pgfplots版本兼容设置
\usepackage{amsmath}          % 数学公式 (acmart通常自带，但加上保险)
\usepackage{booktabs}         % 表格横线 (acmart通常自带)
\usepackage{algorithm}
\usepackage{algorithmic}
\usepackage{float}      % 备用：提供 [H]
\usepackage{placeins}   % 提供 \FloatBarrier
\usepackage{multirow}
\usepackage[table]{xcolor}
\usepackage{tabularx}

\begin{document}

\title{NutriOrion: A Hierarchical Multi-Agent Framework for Personalized Nutrition Intervention Grounded in Clinical Guidelines}


\author{Junwei Wu}
\orcid{1234-5678-9012}
\email{junwei.wu@emory.edu}
\affiliation{%
  \institution{Emory University}
  \city{Atlanta}
  \state{GA}
  \country{USA}
}

\author{Runze Yan}
\orcid{1234-5678-9012}
\email{runze.yan@emory.edu}
\affiliation{%
  \institution{Emory University}
  \city{Atlanta}
  \state{GA}
  \country{USA}
}

\author{Hanqi Luo}
\orcid{1234-5678-9012}
\email{hanqi.luo@emory.edu}
\affiliation{%
  \institution{Emory University}
  \city{Atlanta}
  \state{GA}
  \country{USA}
}

\author{Darren Liu}
\orcid{1234-5678-9012}
\email{darren.liu@emory.edu}
\affiliation{%
  \institution{Emory University}
  \city{Atlanta}
  \state{GA}
  \country{USA}
}

\author{Minxiao Wang}
\orcid{1234-5678-9012}
\email{minxiao.wang@emory.edu}
\affiliation{%
  \institution{Emory University}
  \city{Atlanta}
  \state{GA}
  \country{USA}
}

\author{Kimberly L. Townsend}
\orcid{1234-5678-9012}
\email{ktownsend3@twu.edu}
\affiliation{%
  \institution{Texas Woman's University}
  \city{Denton}
  \state{TX}
  \country{USA}
}

\author{Lydia S. Hartwig}
\orcid{1234-5678-9012}
\email{lrschinn@gmail.com}
\affiliation{%
  \institution{Cecelia Health}
  \city{New York}
  \state{NY}
  \country{USA}
}

\author{Derek Milketinas}
\orcid{1234-5678-9012}
\email{dmiketinas@twu.edu}
\affiliation{%
  \institution{Texas Woman's University}
  \city{Denton}
  \state{TX}
  \country{USA}
}

\author{Xiao Hu}
\orcid{1234-5678-9012}
\email{xiao.hu@emory.edu}
\affiliation{%
  \institution{Emory University}
  \city{Atlanta}
  \state{GA}
  \country{USA}
}

\author{Carl Yang}
\orcid{1234-5678-9012}
\email{j.carlyang@emory.edu}
\affiliation{%
  \institution{Emory University}
  \city{Atlanta}
  \state{GA}
  \country{USA}
}

\begin{abstract}
Personalized nutrition intervention for patients with multimorbidity is critical for improving health outcomes, yet remains challenging as it requires the simultaneous integration of heterogeneous clinical conditions, medications, and dietary guidelines. Single-agent Large Language Models (LLMs) often suffer from context overload and attention dilution when processing such high-dimensional patient profiles. We introduce \textbf{NutriOrion}, a hierarchical multi-agent framework featuring a Parallel-then-Sequential reasoning topology. NutriOrion decomposes nutrition planning into specialized domain agents with isolated contexts to mitigate anchoring bias, followed by a conditional refinement stage. The framework incorporates a multi-objective prioritization algorithm to resolve conflicting dietary requirements and a Safety Constraint Mechanism that injects pharmacological contraindications as hard negative constraints during synthesis, ensuring clinical validity by construction rather than post-hoc filtering. To ensure clinical interoperability, NutriOrion maps synthesized insights into the ADIME standard and FHIR R4 resources. Evaluated on 330 stroke patients with multimorbidity, NutriOrion significantly outperforms multiple baselines, including GPT-4.1 and alternative multi-agent architectures. It achieves a 12.1\% drug-food interaction violation rate, demonstrates strong personalization with negative correlations ($-0.26$ to $-0.35$) between patient biomarkers and recommended risk nutrients, and produces clinically meaningful improvements, including a 167\% increase in fiber and 27\% increase in potassium, alongside reductions in sodium (9\%) and sugars (12\%).
\end{abstract}






%   \centering
%   \includegraphics[width=\textwidth]{NutriOrion26/figure/fig1.pdf} 
%   \caption{Conceptual comparison of reasoning paradigms for high-stakes decision making. }  \description{(a) Monolithic LLM Approach: Standard single-agent models process heterogeneous data (biomarkers, guidelines, medications) in a unified context window. This often leads to information entanglement, where conflicting constraints are hallucianted or ignored due to attention dilution. (b) \textbf{NutriOrion} (Ours): A hierarchical modular framework that decouples reasoning into specialized streams. By isolating domain contexts (e.g., pharmacological constraints vs. nutritional goals) and enforcing deterministic safety gates, NutriOrion synthesizes complex inputs into precise, verifiable, and structured interventions without reasoning collapse.}

\maketitle

\vspace{-0.5em}

\vspace{-0.5em}

\section{Introduction}

\begin{figure*}[t]
  \centering
  \includegraphics[width=\textwidth]{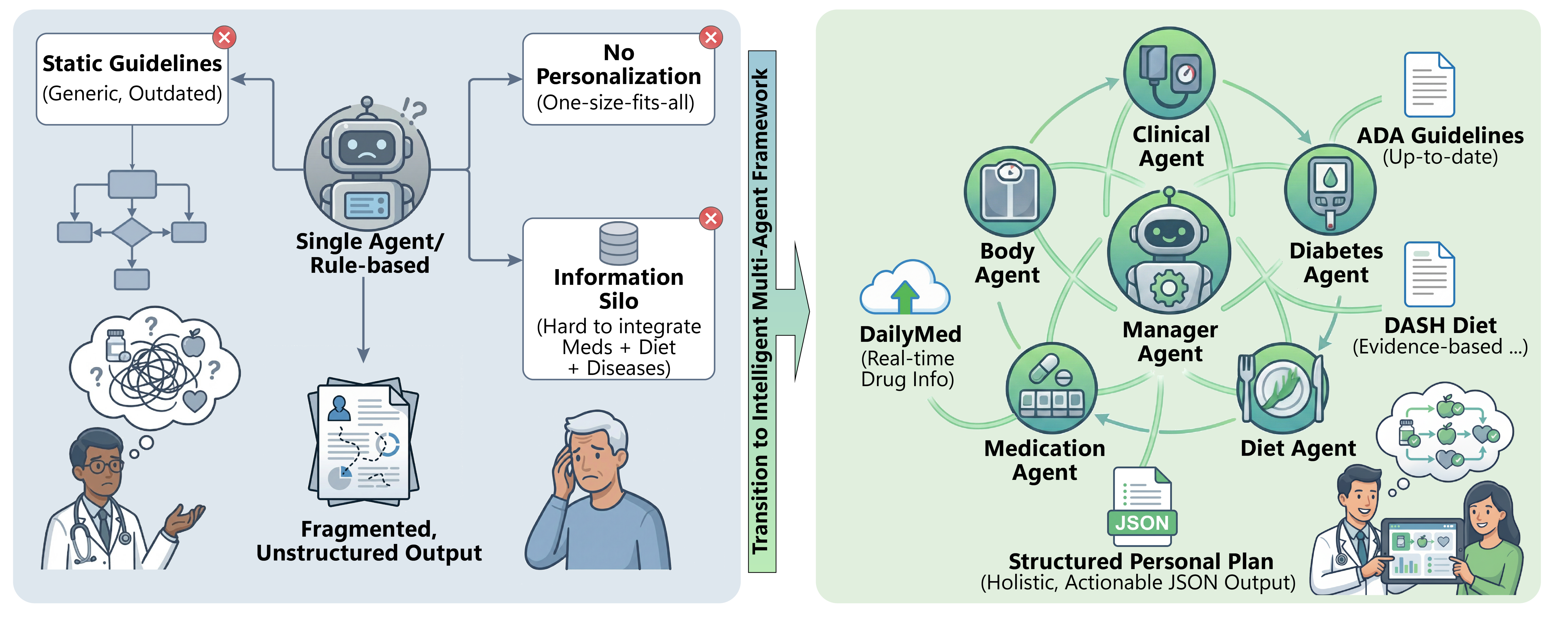} 
  \caption{Conceptual comparison of reasoning paradigms for high-stakes decision making. (a) Monolithic LLM Approach: Standard single-agent models process heterogeneous data in a unified context window. This often leads to information entanglement, where conflicting constraints are hallucinated or ignored due to attention dilution. (b) \textbf{NutriOrion} (Ours): A hierarchical modular framework that decouples reasoning into specialized streams. By isolating domain contexts and enforcing deterministic safety gates, NutriOrion synthesizes complex inputs into precise, verifiable, and structured interventions without reasoning collapse.}
  \label{fig:teaser}
\end{figure*}

Dietary intervention is critical for managing chronic diseases, which account for over 70\% of global mortality~\cite{ELKOMY2024102043}. While structured nutrition can achieve therapeutic effects comparable to pharmacological treatments~\cite{diabetes2002reduction, estruch2018primary}, personalization is hindered by multimorbidity, affecting 40\% of chronic disease patients, where co-occurring conditions often demand conflicting dietary strategies~\cite{schiltz2022prevalence, chowdhury2023global}. Furthermore, a severe global shortage of nutrition professionals (1:5,000 ratio) makes manual integration of patient data with complex clinical guidelines nearly impossible at scale~\cite{ICDA_WorkforceReport2021, siopis2020dietetic}.

Recent years have witnessed growing research in computational dietary recommendation systems, with several distinct methodological approaches emerging~\cite{min2019survey, trang2018overview}. Early systems employed collaborative filtering and content-based methods, recommending foods by matching user preferences with similar users or item attributes~\cite{elsweiler2017exploiting, kalpakoglou2025ai, amiri2024delighting}. A second line of work developed constraint-based approaches that explicitly encode dietary restrictions as hard rules, filtering out foods that violate specific nutritional constraints such as sodium limits for hypertension management~\cite{kalpakoglou2025ai, ribeiro2017souschef}. In contrast, optimization-based methods formulate meal planning as mathematical programming problems, where algorithms search for meal combinations that simultaneously optimize multiple nutritional objectives under various constraints~\cite{padovan2023optimized, gazan2018mathematical}. More recent hybrid approaches combine rule-based medical knowledge with machine learning models, using expert-defined nutritional guidelines to constrain the recommendation space while allowing data-driven learning to adapt to individual user preferences and dietary patterns~\cite{agapito2018dietos, papastratis2024ai}. The latest deep learning methods learn representations by embedding foods and users into continuous vector spaces, enabling prediction of health outcomes and personalized recommendations based on individual profiles and contextual factors~\cite{rostami2022effective, metwally2021learning}.

% \begin{figure*}
%   \centering
%   \includegraphics[width=\textwidth]{NutriOrion26'/figure/DiagramFinal-03.pdf} 
%   \caption{Conceptual comparison of reasoning paradigms for high-stakes decision making. (a) Monolithic LLM Approach: Standard single-agent models process heterogeneous data in a unified context window. This often leads to information entanglement, where conflicting constraints are hallucinated or ignored due to attention dilution. (b) \textbf{NutriOrion} (Ours): A hierarchical modular framework that decouples reasoning into specialized streams. By isolating domain contexts and enforcing deterministic safety gates, NutriOrion synthesizes complex inputs into precise, verifiable, and structured interventions without reasoning collapse.}
% \end{figure*}

However, despite these methodological advances in computational dietary recommendation, existing systems remain inadequate for clinical deployment due to four fundamental limitations. First, they optimize for generic health scores rather than disease-specific clinical objectives defined in existing dietary guidelines. Most systems use composite wellness metrics or broad nutritional balance scores without operationalizing the precise quantitative targets specified in evidence-based protocols for managing specific chronic diseases. Second, they fail to provide personalized modifications to actual dietary intake. Instead of analyzing what patients currently eat and suggesting actionable adjustments, most systems generate idealized meal plans disconnected from real-world eating behaviors, making adherence difficult. Third, they cannot handle multimorbidity scenarios with conflicting dietary requirements. Existing approaches either process diseases independently (generating incompatible recommendations) or aggregate objectives into a single function (losing clinical prioritization logic). Fourth, they lack robust verification of medication-diet interactions and strict adherence to existing dietary guidelines for each comorbidity. With multiple medications and diseases, existing systems do not perform systematic cross-domain verification for safety across pharmacological constraints~\cite{patel2016evaluation, anandabaskar2019drug}.

Large Language Models (LLMs) offer new possibilities for addressing these challenges through their capabilities in multi-domain reasoning and integrating heterogeneous information~\cite{shool2025systematic, yang2025behavioral}. However, existing work shows that monolithic LLM architectures suffer from reasoning collapse when processing high-dimensional feature spaces~\cite{liu2024lost, qiu2025quantifying}. In personalized nutrition for multimorbidity, single-agent models face this challenge when processing comprehensive patient profiles containing dozens of heterogeneous features: demographics, anthropometrics, laboratory biomarkers, medication lists, dietary histories, and existing dietary guidelines, as illustrated in Figure~\ref{fig:teaser}. As context complexity increases, models generate recommendations violating safety constraints, hallucinate interventions contradicting patient data, and exhibit attention dilution where important constraints get overridden~\cite{asgari2025framework, roustan2025clinicians}. This occurs because monolithic architectures process all information through unified attention, causing cross-domain interference when reasoning spans multiple knowledge domains~\cite{liu2024lost, borkowski2025multiagent}. Furthermore, probabilistic safety mechanisms (RLHF, constitutional AI) cannot provide deterministic guarantees required for medical decision support where rare failures cause serious adverse events~\cite{asgari2025framework, roustan2025clinicians}.

We propose NutriOrion, a hierarchical multi-agent framework that modularizes nutrition intervention into a Parallel-then-Sequential reasoning topology. In Stage 1, domain-specific agents (Body, Clinical, Medication, and Diet) maintain isolated contexts to perform independent assessment grounded in dual-source retrieval from clinical guidelines and pharmacological databases \cite{lewis2020retrieval}. This independence mitigates anchoring bias and prevents cross-domain hallucination. In Stage 2, specialized dietitians perform conditional refinement, using diagnostic insights from Stage 1 as reasoning constraints. To resolve multimorbidity conflicts, a Health Prioritization Agent performs synthesis using multi-objective scoring $S(j)$ based on severity, urgency, and modifiability. Crucially, rather than relying on post-hoc filtering, NutriOrion implements a Safety Constraint Mechanism that extracts pharmacological contraindications and injects them as hard negative constraints during final generation, ensuring clinical validity by construction. Finally, the framework projects synthesized insights into the ADIME standard \cite{skipper2007applying} and performs an isomorphic mapping $\Phi$ to FHIR R4 resources, ensuring seamless EHR interoperability.

\textbf{Contributions}. This work makes four primary contributions: (1) a hierarchical multi-agent architecture with isolated contexts and domain-specific RAG that prevents reasoning collapse, grounds recommendations in existing dietary guidelines for disease-specific clinical objectives, and enables personalized modifications to actual dietary intake; 
(2) a Severity-Urgency-Modifiability scoring mechanism enabling dynamic, patient-specific guideline conflict resolution in multimorbidity scenarios; 
(3) a hybrid architecture separating probabilistic generation from deterministic verification against medication-diet contraindication databases, providing low-failure-rate safety guarantees; and 
(4) comprehensive evaluation on 330 multimorbid stroke patients from the the National Health and Nutrition Examination Survey (NHANES) dataset~\cite{johnson2014national} across five complementary dimensions: output structure and actionability, medication safety, clinical appropriateness, personalization, and food quality, demonstrating that NutriOrion significantly outperforms strong baselines including GPT-4.1 and Claude-Sonnet-4 with various prompting strategies and alternative multi-agent architectures.

\vspace{-1em}

% This work makes four primary contributions to the field of AI for healthcare:

% \begin{itemize}
%     \item \textbf{Hierarchical Reasoning Framework}: We formalize nutritional intervention as a composite function where specialized agents execute parallel domain analysis followed by sequential refinement. This modular architecture prevents reasoning collapse by isolating context, enabling deep reasoning on high-dimensional electronic health records.
    
%     \item \textbf{Multi-Objective Conflict Resolution}: We introduce a mathematical prioritization algorithm that weights identified health issues by Severity, Urgency, and Modifiability. This mechanism allows the system to dynamically resolve the ``Guideline Paradox'' (e.g., prioritizing hypoglycemia risk over long-term cholesterol goals).
    
%     \item \textbf{Deterministic Safety Verification}: Distinct from standard Reinforcement Learning with Human Feedback (RLHF) guardrails, we implement a post-hoc Safety Gate that rigidly filters interventions against structured contraindications extracted from federal databases (DailyMed), ensuring a ``Do No Harm'' standard.
    
%     \item \textbf{Rigorous Empirical Validation}: We evaluate NutriOrion on a diverse cohort of 330 multimorbid patients derived from the National Health and Nutrition Examination Survey (NHANES). Results demonstrate that our framework achieves superior evidence grounding and zero safety violations, significantly outperforming strong single-agent baselines in both clinical expert scoring and structural completeness.
% \end{itemize}
\section{Related Work} \label{sec:related_work}

\subsection{LLM Agents in Healthcare}
Large Language Models (LLMs) have demonstrated significant potential in clinical tasks, ranging from medical question-answering \cite{singhal2023large} to generating clinical notes \cite{yang2023large}. Recent work has explored deploying multi-agent systems to mimic clinical workflows and improve reasoning capabilities \cite{tang2024medagents, wang2024survey}. For instance, MedAgents \cite{tang2024medagents} utilizes a collaborative multi-agent structure to enhance diagnosis accuracy. However, existing approaches often rely on general medical knowledge and lack specific mechanisms for handling complex, multi-modal patient data (NHANES) and resolving multi-morbidity conflicts in nutrition intervention. NutriOrion addresses this by introducing a specialized parallel-then-sequential topology with explicit bias mitigation and constraint propagation mechanisms, designed specifically for personalized nutrition planning.

\subsection{Retrieval-Augmented Grounding and Safety}
Retrieval-Augmented Generation (RAG) has emerged as a critical technique to mitigate LLM hallucination by grounding generation in external knowledge bases \cite{lewis2020retrieval, gao2023retrieval}. In the medical domain, RAG has been applied to ground responses in verified clinical guidelines and biomedical literature \cite{zakka2024almanac, chen2024medrag}. While effective for general information retrieval, prior work often lacks robust mechanisms to enforce strict clinical safety constraints, particularly regarding drug-nutrient interactions (DNIs). NutriOrion advances this by integrating a dual-source retrieval system that accesses both clinical guidelines (DASH, ADA) and structured pharmacological data (DailyMed). Crucially, we implement a retrieval-augmented safety constraint mechanism that treats retrieved DNI information as hard negative constraints during synthesis, ensuring output clinical validity by construction.

\vspace{-0.5em}

\vspace{-0.5em}

\section{Methodology} \label{sec:methodology}

\begin{figure*}
    \centering
    \includegraphics[width=\linewidth]{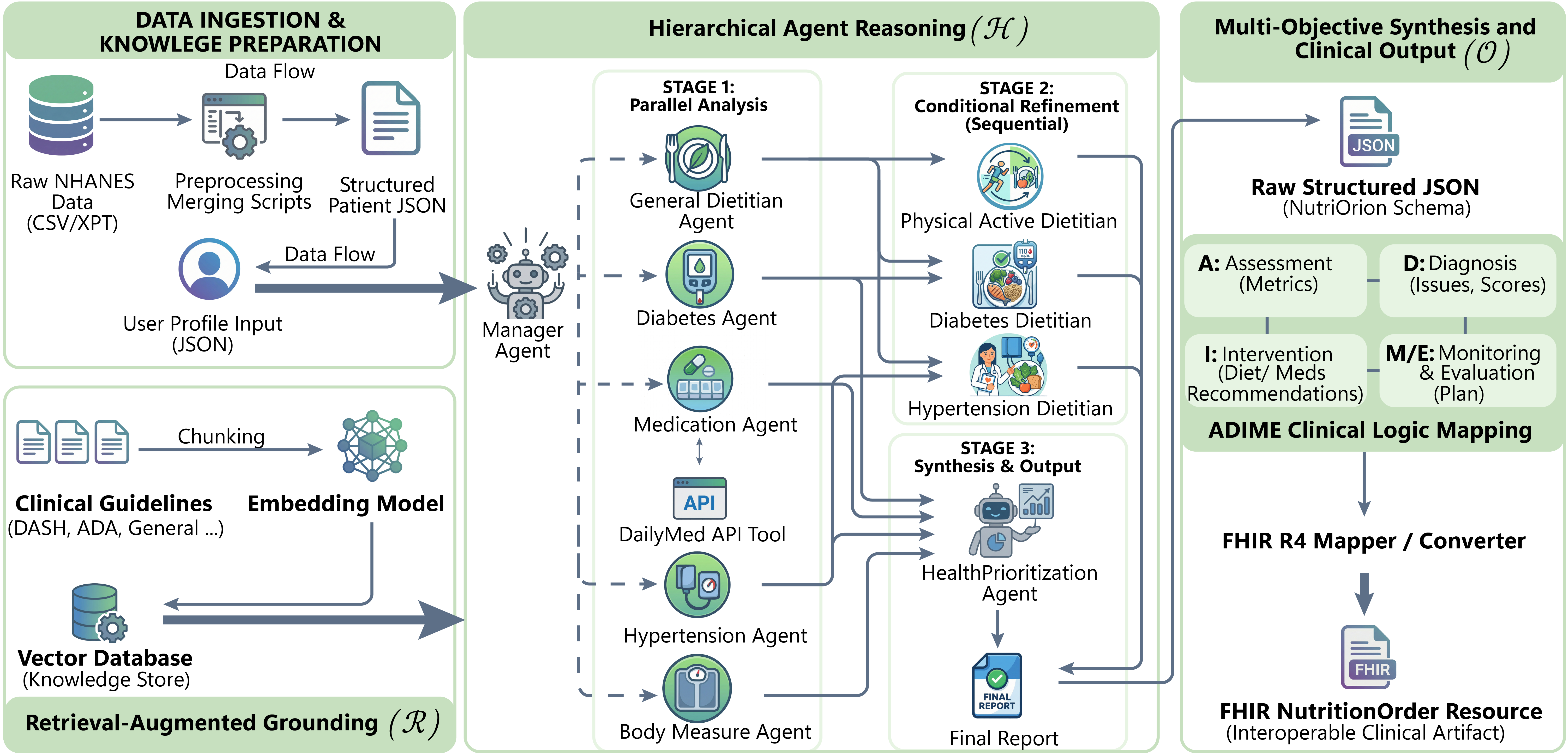}
\caption{\textbf{The NutriOrion System Pipeline.} The framework integrates three modular components: 
  \textbf{(Left) Retrieval-Augmented Grounding ($\mathcal{R}$):} Ingestion of patient profiles and vectorization of clinical guidelines (e.g., DASH, ADA); 
  \textbf{(Center) Hierarchical Agent Reasoning ($\mathcal{H}$):} A three-stage orchestration consisting of parallel diagnostic analysis (Stage 1), sequential dietary refinement (Stage 2), and safety-aware synthesis (Stage 3); 
  \textbf{(Right) Output Standardization ($\mathcal{O}$):} Projection of synthesized ADIME-structured insights into interoperable FHIR R4 clinical resources.}
    \label{fig:pip}
\end{figure*}

We formulate the problem of personalized nutrition intervention as a structured inference task mapping a multi-modal patient state $X$ to a clinically valid action plan $Y$. To achieve this, we propose \textbf{NutriOrion}, a framework shown in Figure~\ref{fig:pip} that integrates hierarchical agentic reasoning with strict evidence grounding.

\subsection{Framework Formulation}

Let $\mathcal{P}$ denote the patient population. For each patient $p \in \mathcal{P}$, the input state $X_p \in \mathcal{X}$ is a composite vector of clinical and demographic features:
\begin{equation}
X_p = [A_p, B_p, M_p, D_p, S_p]
\end{equation}
where $A_p$ represents anthropometrics (e.g., BMI), $B_p$ clinical biomarkers (e.g., HbA1c, BP), $M_p$ the medication profile, $D_p$ dietary history, and $S_p$ sociodemographic factors.

The objective is to learn a mapping function $F: \mathcal{X} \to \mathcal{Y}$ that generates an output $Y_p$ strictly adhering to the ADIME (Assessment, Diagnosis, Intervention, Monitoring, Evaluation) clinical standard. We decompose $F$ into a hierarchical composition of specialized agent functions $f_i$, grounded by a retrieval function $\mathcal{R}(\cdot)$:
\begin{equation}
Y_p = F(X_p; \mathcal{K}, \Theta) = \mathcal{O} \circ \mathcal{H}(X_p, \mathcal{R}(X_p, \mathcal{K}))
\end{equation}
Here, $\mathcal{K}$ represents the external knowledge base, $\mathcal{H}$ is the hierarchical reasoning engine, and $\mathcal{O}$ is the output synthesis function. The system parameters $\Theta$ (LLM weights) remain fixed with temperature $T=0$ to ensure deterministic execution.

\subsection{Retrieval-Augmented Grounding ($\mathcal{R}$)}

To ensure clinical validity and mitigate hallucination, we construct a dual-source retrieval system $\mathcal{R}$ that maps agent queries $q$ to evidence contexts $C$.

\subsubsection{Guideline Knowledge Manifold}
We define a vector space $\mathcal{V} \subset \mathbb{R}^d$ populated by embeddings of verified clinical guidelines (DASH, ADA, USDA). Let $D_{guide}$ be the set of guideline documents. We apply a chunking function $\phi: D_{guide} \to \{c_1, \dots, c_m\}$ to segment documents into semantic units. Each chunk $c_j$ is mapped to an embedding vector $\mathbf{e}_j \in \mathbb{R}^d$ using the BGE-M3 model.

For a given query $q$, the retrieval function $r_{guide}(q)$ returns the top-$k$ relevant chunks based on cosine similarity:
\begin{equation}
r_{guide}(q) = \underset{C' \subset \{c_j\}, |C'|=k}{\arg\max} \sum_{c \in C'} \frac{\mathbf{e}_q \cdot \mathbf{e}_c}{\|\mathbf{e}_q\| \|\mathbf{e}_c\|}
\end{equation}
where $k=3$ is selected to balance context window constraints with evidence sufficiency.

\subsubsection{Structured Tool Retrieval}
For medication analysis, we define a discrete retrieval function $r_{drug}(M_p)$ that queries the DailyMed/openFDA database. This function maps a drug set $M_p$ to a set of Structured Product Labels (SPLs), filtering for sections $S_{food} \subset SPL$ related to drug-nutrient interactions.

\vspace{-0.5em}

\vspace{-0.5em}

\subsection{Hierarchical Agent Reasoning ($\mathcal{H}$)}

The inference process is modeled as a directed execution graph $\mathcal{G}_{exec} = (\mathcal{A}, E)$. The agent set is defined as $\mathcal{A} = \{a_{mgr}\} \cup \mathcal{A}_{domain} \cup \mathcal{A}_{refine} \cup \mathcal{A}_{synth}$. Each agent $a_i$ is parameterized by a Large Language Model $\mathcal{M}$ and a specific role prompt $\rho_i$.

\subsubsection{Execution Topology \& Rationale}
We design the execution topology to mirror the clinical decision-making process for multimorbidity management. We adopt a Parallel-then-Sequential architecture (see Algorithm \ref{alg:fim_agent}) grounded in three clinical principles:

\begin{itemize}
    \item \textbf{Bias Mitigation via Independence (Parallel Analysis)}: In Stage 1, domain agents ($\mathcal{A}_{domain}$) process specific subspaces of the patient profile $X_p$ concurrently. This isolation ensures that the assessment of one condition (e.g., Diabetes) is not biased by the preliminary findings of another (e.g., Hypertension). This design simulates the \textit{independent assessment} phase of specialized clinicians, effectively preventing the propagation of hallucinations or anchoring bias across domains.
    
    \item \textbf{Constraint Propagation (Sequential Refinement)}: In Stage 2, condition-specific dietitians ($\mathcal{A}_{refine}$) refine the general dietary assessment based on the \textit{confirmed} diagnoses from Stage 1. This models the clinical workflow of \textit{"Consultation \& Adjustment,"} where a diagnosis acts as a conditional constraint on subsequent lifestyle interventions (e.g., a confirmed hypertension diagnosis activates DASH-specific sodium limits).
    
    \item \textbf{Multi-Objective Optimization (Synthesis)}: The final stage addresses the "multi-morbidity conflict" problem. By aggregating all upstream constraints, the system performs a synthesis that balances conflicting guidelines (e.g., renal vs. diabetic diets), effectively solving a multi-objective optimization problem to maximize patient health utility under safety constraints.

\end{itemize}

\subsubsection{Algorithmic Flow}
The formal inference procedure is detailed in Algorithm \ref{alg:fim_agent}. Crucially, the process integrates both soft reasoning (LLM-based) and hard safety checks via contextual constraint injection.

\begin{algorithm}[t]
\caption{NutriOrion Inference Process}
\label{alg:fim_agent}
\begin{algorithmic}[1]
\REQUIRE Patient Profile $X_p$, Guidelines $\mathcal{K}$
\ENSURE Personalized Plan $Y_p$
\STATE \textbf{Initialize} context $\mathcal{C} \leftarrow \emptyset$

\STATE \COMMENT{\textbf{Stage 1: Independent Assessment (Parallel)}}
\FOR{agent $a_i \in \mathcal{A}_{domain}$}
    \STATE Query $q_i \leftarrow a_i.\text{draft\_query}(X_p)$
    \STATE Evidence $E_i \leftarrow \mathcal{R}(q_i, \mathcal{K})$
    \STATE Insight $h_i \leftarrow a_i.\text{analyze}(X_p, E_i)$
    \STATE $\mathcal{C} \leftarrow \mathcal{C} \cup \{h_i\}$
\ENDFOR

\STATE \COMMENT{\textbf{Stage 2: Conditional Refinement (Sequential)}}
\FOR{specialist $a_{spec} \in \mathcal{A}_{refine}$}
    \STATE Context $C_{spec} \leftarrow \text{SelectContext}(\mathcal{C}, a_{spec})$
    \STATE $r_{spec} \leftarrow a_{spec}.\text{reason}(D_p, C_{spec}, \mathcal{K}_{spec})$
    \STATE $\mathcal{C} \leftarrow \mathcal{C} \cup \{r_{spec}\}$
\ENDFOR

\STATE \COMMENT{\textbf{Stage 3: Synthesis \& Safety Verification}}
\STATE Priority Vector $V_{prio} \leftarrow a_{prio}(\mathcal{C})$
\STATE Ranked Issues $L \leftarrow \text{Rank}(V_{prio})$ using Eq. (\ref{eq:scoring})

\STATE \COMMENT{Extract Hard DNI Constraints from Medication Agent}
\STATE $C_{safe} \leftarrow \text{ExtractConstraints}(\mathcal{C}, a_{med})$

\STATE \COMMENT{Generate plan strictly bounded by safety context}
\STATE $Y_p \leftarrow a_{report}(L, \mathcal{C}, C_{safe})$

\RETURN $Y_p$
\end{algorithmic}
\end{algorithm}

The Safety Constraint Mechanism (Algorithm \ref{alg:fim_agent}, lines 17-19) ensures pharmacological safety. Unlike post-hoc filtering which risks plan incoherence, this stage extracts contraindications ($C_{safe}$) from Stage 1 and injects them as hard negative constraints during the final synthesis. This guarantees that the generated plan $Y_p$ is chemically valid by construction.

\subsection{Multi-Objective Synthesis and Clinical Output ($\mathcal{O}$)}

The synthesis stage aggregates the disparate insights from domain agents into a cohesive clinical plan. This process is governed by a Health Prioritization Agent and a Structured Projection Module.

\subsubsection{Priority Scoring Mechanism}
To resolve conflicting guidelines (e.g., DASH requiring high potassium vs. CKD requiring restriction), we define a scoring function $S(j)$ for each identified health issue $j$:
\begin{equation}
\label{eq:scoring}
S(j) = w_{sev} \cdot \sigma_{sev}(j) + w_{urg} \cdot \sigma_{urg}(j) + w_{mod} \cdot \sigma_{mod}(j)
\end{equation}
where $\sigma \in [0, 1]$ represents normalized scores derived from clinical thresholds for Severity, Urgency, and Modifiability. Issues are ranked by $S(j)$ to determine the primary intervention focus.

\subsubsection{ADIME-Structured Projection}
Unlike generic LLMs that generate free-text narratives, NutriOrion projects the synthesized insights into a strict ADIME (Assessment, Diagnosis, Intervention, Monitoring, Evaluation) schema. The intervention component of the plan, $I_{plan} \subset Y_p$, is structured as discrete, actionable directives:

\begin{equation}
\resizebox{0.95\linewidth}{!}{$
    I_{plan} = \{ (action, food, reason) \mid action \in \{\texttt{Continue}, \texttt{Replace}, \texttt{Add}\} \}
$}
\end{equation}

This structured classification—explicitly separates positive reinforcement (\texttt{Continue}) from corrective substitutions (\texttt{Replace})—aligns with the standardized terminology of the Nutrition Care Process (NCP).

\subsubsection{Interoperability: FHIR-Aligned Output Projection}
To bridge the gap between AI-driven generation and clinical deployment, we construct the output space $\mathcal{Y}$ to be formally isomorphic to the \textbf{HL7 FHIR R4 \texttt{NutritionOrder}} resource. We define a deterministic mapping function $\Phi: Y_p \to Y_{FHIR}$ that transforms the finalized ADIME components into standardized, EHR-conformant artifacts. Specifically, agent-generated dietary protocols are projected onto standardized clinical terminologies, while quantitative constraints—such as upper bounds on sodium intake—are encoded as structured nutrient definitions. A comprehensive specification is provided in Appendix \ref{sec:fhir_mapping}. By design, this architectural alignment renders NutriOrion inherently ``FHIR-ready,'' enabling seamless integration into hospital information systems.

\subsection{Safety Verification Mechanism ($\mathcal{G}_{safe}$)}
\label{sec:safety_vefi_mech}

While hierarchical synthesis optimizes for clinical utility, patient safety requires strict adherence to pharmacological contraindications. Instead of relying on probabilistic knowledge, we implement a Retrieval-Augmented Safety Constraint mechanism.

Let $C_{safe} \subset \mathcal{C}$ be the set of Drug-Nutrient Interactions (DNIs) identified by the Medication Agent using the DailyMed tool (Stage 1). During the final synthesis (Stage 3), these interactions are treated not as optional context, but as Hard Negative Constraints in the inference prompt. Formally, the synthesis function maximizes clinical utility $U(Y_p)$ subject to:
\begin{equation}
\forall d \in C_{safe}, \quad \text{Sim}(Y_p, d_{contraindication}) < \epsilon
\end{equation}

\vspace{-0.3cm}
\section{Experiments}\label{sec:exp}

We evaluate NutriOrion across five complementary dimensions: \textbf{output structure and actionability} (Section~\ref{sec:doc-quality}) measures adherence to standardized care formats and recommendation specificity, \textbf{medication safety} (Section~\ref{sec:safety-audit}) ensures interventions avoid drug-nutrient interactions, \textbf{clinical appropriateness} (Section~\ref{sec:clinical_appr}) assesses alignment with evidence-based guidelines, \textbf{personalization} (Section~\ref{sec:personalization}) quantifies biomarker-driven dietary tailoring, and \textbf{food quality} (Section~\ref{sec:food_quality}) evaluates nutritional healthfulness and compositional improvements. These dimensions collectively validate clinical deployment readiness.

\subsection{Experimental Setup}

\subsubsection{Dataset and Patient Cohort}

We evaluated all 330 stroke survivors from the NHANES 2011-2012 cycles~\cite{johnson2014national}. Stroke survivors frequently present with multimorbidities, particularly hypertension and diabetes, making them ideal for evaluating personalized nutrition intervention in complex clinical scenarios~\cite{yan2023association}. All patients had complete data required by NutriOrion, including dietary recall (24-hour recall), laboratory biomarkers (e.g., HbA1c, blood pressure, lipid panel), and medication records.

\subsubsection{Baselines and Ablation Studies}
\label{sec:exp_setup}

To rigorously evaluate NutriOrion(using GPT4o-mini), we benchmark against 12 baselines sharing identical tool access and output schemas. We categorize comparisons into three research questions (RQ).

\noindent\textbf{RQ1: Can Single Agents Replace Multi-Agent Architectures?} We investigate whether a single powerful LLM can match multi-agent precision by evaluating three reasoning paradigms on open-weights models (Llama-3.1-70B, Qwen-2.5-72B): \textbf{ReAct} \cite{yao2022react} (standard dynamic thought-action loop), \textbf{Chain-of-Thought (CoT)} \cite{wei2022chain} (linear step-by-step reasoning without backtracking), and \textbf{Self-Refine} \cite{madaan2023self} (iterative \textit{Draft-Critique-Refine} pipeline). Additionally, we benchmark against closed-source models (\textbf{GPT-4.1}, \textbf{Claude-4-Sonnet}) using ReAct to determine if foundation model capability overrides architectural benefits.

\noindent\textbf{RQ2: How Do Roles and Process Contribute?} To disentangle NutriOrion's core components, we conduct controlled ablations: \textbf{w/o ADIME Structure} retains 10 specialized agents but removes the DAG dependency to rely on dynamic manager orchestration; \textbf{w/o Specialized Roles} retains the ADIME workflow but replaces domain experts with generic "Clinical Nutritionist" agents to test the necessity of domain-specific backstories.

\noindent\textbf{RQ3: Is Structured Collaboration Superior?} We compare against two prevalent multi-agent paradigms: \textbf{Unstructured Collaboration (Round Table)} simulating \textbf{AutoGen}-style \cite{wu2023autogen} free discussion without a predefined pipeline, and \textbf{Ensemble Voting} simulating \textbf{MedAgents}-style \cite{tang2023medagents} redundancy where parallel agents with diverse personas (Conservative, Proactive, Evidence-Based) vote on recommendations. Table \ref{tab:baselines} summarizes all methods. Implementation details are provided in Appendix \ref{sec:appendix_baselines}.

\vspace{-1.3em}

% --- Summary Table ---
% \begin{table}[h]
% \centering
% \caption{Summary of Baselines. All methods use temperature=0.0 and identical tools.}
% \label{tab:baselines}
% \resizebox{\columnwidth}{!}{%
% \begin{tabular}{lcccc}
% \toprule
% \textbf{Method} & \textbf{Type} & \textbf{Logic} & \textbf{\# Agents} & \textbf{Core Test} \\
% \midrule
% \textbf{NutriOrion (Ours)} & Multi-Agent & \textbf{Structured ADIME + Specialized Roles} & 10 & SOTA Performance \\
% \midrule
% ReAct \cite{yao2022react} & Single & Dynamic Tool Loop & 1 & Dynamic Reasoning \\
% CoT \cite{wei2022chain} & Single & Linear Planning & 1 & Linear Reasoning \\
% Self-Refine \cite{madaan2023self} & Single & Iterative Correction & 1 & Self-Correction \\
% API ReAct  & Single & Dynamic Tool Loop & 1 & Model Capability \\
% \midrule
% w/o ADIME & Ablation & Dynamic Delegation & 10 & Workflow Impact \\
% w/o Roles & Ablation & Generic Roles + Fixed DAG & 10 & Persona Impact \\
% \midrule
% Round Table \cite{wu2023autogen} & Benchmark & Unstructured Discussion & 3 & Structure vs. Freedom \\
% Ensemble \cite{tang2023medagents} & Benchmark & Parallel Persona Voting & 3+1 & Specialization vs. Redundancy \\
% \bottomrule
% \end{tabular}%
% }
% \end{table}

\begin{table}[h]
\centering
\caption{Summary of baseline methods and ablations. All methods use identical tools and temperature = 0.0.}
\label{tab:baselines}
\resizebox{\columnwidth}{!}{%
\begin{tabular}{lcccc}
\toprule
\textbf{Method} & \textbf{Type} & \textbf{Logic} & \textbf{\# Agents} & \textbf{Core Test} \\
\midrule

\multirow{2}{*}{\textbf{NutriOrion (Ours)}} 
& \multirow{2}{*}{Multi-Agent} 
& \textbf{Structured ADIME} 
& \multirow{2}{*}{10} 
& \multirow{2}{*}{SOTA Performance} \\
& & \textbf{+ Specialized Roles} & & \\
\midrule

ReAct \cite{yao2022react} 
& Single 
& Dynamic Tool Loop 
& 1 
& Dynamic Reasoning \\

CoT \cite{wei2022chain} 
& Single 
& Linear Planning 
& 1 
& Linear Reasoning \\

Self-Refine \cite{madaan2023self} 
& Single 
& Iterative Correction 
& 1 
& Self-Correction \\

API ReAct 
& Single 
& Dynamic Tool Loop 
& 1 
& Model Capability \\
\midrule

w/o ADIME 
& Ablation 
& Dynamic Delegation 
& 10 
& Workflow Impact \\

w/o Roles 
& Ablation 
& Generic Roles + Fixed DAG 
& 10 
& Persona Impact \\
\midrule

Round Table \cite{wu2023autogen} 
& Benchmark 
& Unstructured Discussion 
& 3 
& Structure vs.\ Freedom \\

Ensemble \cite{tang2023medagents} 
& Benchmark 
& Parallel Persona Voting 
& 3+1 
& Specialization vs.\ Redundancy \\
\bottomrule
\end{tabular}%
}
\end{table}

\subsection{Output Structure \& Actionability}
\label{sec:doc-quality}

To ensure clinical validity beyond lexical metrics, we conducted expert evaluation using two Board-Certified Registered Dietitians (RDs). We employed the Nutrition Care Process Quality Evaluation and Standardization Tool (NCP-QUEST)~\cite{lewis2022ncpquest}, a validated instrument for assessing nutrition care documentation quality across the ADIME framework: Assessment, Diagnosis, Intervention, and Monitoring/Evaluation. Evaluators were blinded to model sources and independently scored each plan on NCP-QUEST's 5-point Likert scales, referencing evidence-based dietary guidelines including American Diabetes Association (ADA) standards and Dietary Approaches to Stop Hypertension (DASH) protocols~\cite{ada,dash} to judge clinical appropriateness. Inter-rater agreement was substantial (Cohen's $\kappa=0.82$).

Beyond structural completeness, we introduced a second quality metric: Actionability Rate, which quantifies the proportion of recommendations containing specific, actionable food items versus generic dietary categories. For each plan, RDs annotated recommendations as either specific (e.g., "grilled salmon," "spinach") or generic (e.g., "healthy foods," "more protein"), and we calculated actionability as 1 minus the proportion of generic recommendations. This metric reveals whether models provide practically implementable guidance or merely abstract dietary principles

Table~\ref{tab:ncpquest} presents the dual-evaluation results. NutriOrion achieves an excellent expert rating (7.5/8) and state-of-the-art actionability (97.8\%), effectively matching frontier commercial APIs like GPT-4.1 (91.2\%) and significantly outperforming open-source baselines. Crucially, the actionability metric reveals a "fluency trap" in baseline models. For instance, while React-Qwen-72B achieves a respectable expert score (6.0/8) due to fluent reasoning, its actionability is critically low (20.8\%). This discrepancy is quantified by a moderate Pearson correlation ($r=0.63, p<0.05$) between metrics, confirming that specificity is a necessary but insufficient condition for quality. Models failing to produce structured ADIME outputs receive zero scores for intervention-specific dimensions despite sometimes generating reasonable diagnoses. NutriOrion's specialized agent architecture successfully bridges the gap between abstract medical knowledge and actionable patient instructions, generating approximately $\approx$7 times less output than frontier models (2,663 vs.\ 17,925 average characters) while maintaining comparable clinical accuracy, making it more efficient for reviewing clinicians.

\vspace{-0.3em}

\begin{table}[h]
\centering
\caption{NCP-QUEST Scores \& Actionability Rates. NCP-QUEST scores (0--8) averaged from two independent RDs ($\kappa=0.82$). \textbf{Act.\%} (Actionability) denotes the success rate of generating concrete, identifiable food items (Calculated as $1 - \text{Non-specific Rate}$). }
\label{tab:ncpquest}
\resizebox{\columnwidth}{!}{%
% 注意：lccccc|c 在Total和Act之间加了一条竖线
\begin{tabular}{@{}lccccc|c@{}}
\toprule
\textbf{Model} & \textbf{Diagnosis} & \textbf{Interv.} & \textbf{Logic} & \textbf{Detail} & \textbf{Total} & \textbf{Act.\%} \\ 
\midrule
Claude-Sonnet-4    & 2.0 & 2.0 & 1.5 & 1.5 & 7.0 & 91.2 \\
GPT-4.1            & 2.0 & 1.5 & 1.5 & 1.5 & 6.5 & 91.2 \\
\textbf{NutriOrion}& \textbf{2.0} & \textbf{2.0} & \textbf{2.0} & \textbf{1.5} & \textbf{7.5} & \textbf{97.8} \\ 
React-Qwen-2.5-72B    & 2.0 & 1.5 & 1.5 & 1.0 & 6.0 & 20.8 \\
React-LLaMA-3-70B     & 1.5 & 1.0 & 1.5 & 0.5 & 4.5 & 18.1 \\ 
\midrule
\multicolumn{7}{l}{\textit{Models below produce non-structured outputs:}} \\
Ablation-NoSpec.      & 2.0 & 0.0 & 0.0 & 0.0 & 2.0 & 85.6 \\
SelfRefine-LLaMA-3-70B  & 1.5 & 0.0 & 0.0 & 0.0 & 1.5 & 43.4 \\
SelfRefine-Qwen2.5-72B   & 1.0 & 0.0 & 0.0 & 0.0 & 1.0 & 32.9 \\
CoT-Qwen-2.5-72B          & 1.0 & 0.0 & 0.0 & 0.0 & 1.0 & 7.8 \\ 
CoT-LLaMA-3-70B         & 1.0 & 0.0 & 0.0 & 0.0 & 1.0 & 26.4 \\
Ablation-NoADIME      & 0.0 & 0.0 & 0.0 & 0.0 & 0.0 & 24.2 \\
Multiagent-RoundTable & 1.0 & 0.0 & 0.0 & 0.0 & 1.0 & 19.5 \\
Multiagent-Ensemble   & 0.0 & 0.0 & 0.0 & 0.0 & 0.0 & 37.6 \\
\bottomrule
\end{tabular}%
}
\end{table}

\vspace{-0.5em}

\vspace{-0.5em}

\subsection{Medication Safety}
\label{sec:safety-audit}

Drug-food interactions represent a critical safety boundary in clinical nutrition, particularly for stroke patients managing polypharmacy~\cite{bushra2011food}. To assess this, we implemented a rigorous safety audit employing a two-stage validation pipeline: (1) static screening against guideline-derived contraindication lists, followed by (2) ground-truth verification against the FDA DailyMed database~\cite{dailymed}, where we queried specific drug labels to extract authoritative warning sections. 

We focused on three high-risk post-stroke cohorts: Warfarin users ($n=28$), Potassium-sparing diuretic users ($n=15$), and Statin users ($n=114$). These groups were selected for their well-documented and potentially severe dietary interactions. The underlying pharmacological mechanisms, contraindicated food items, and corresponding guideline references are detailed in Appendix~\ref{tab:safety-scenario}.

\begin{table}[t]
\centering
\caption{Drug-Food Interaction Violations. \textbf{Eval.} ($N$) denotes the number of outputs containing actionable food items. \textbf{Rate} is the percentage of outputs triggering FDA contraindications. \textcolor{gray}{Rows in gray} indicate models with insufficient Actionability ($<40\%$ in Table~\ref{tab:ncpquest}). Their near-zero violation rates are artifacts of silence (failing to prescribe food) rather than safety.}
\label{tab:safety}
\small 
\begin{tabular*}{\columnwidth}{@{\extracolsep{\fill}}lccc@{}}
\toprule
\textbf{Model} & \textbf{Eval. ($N$)} & \textbf{Violation} & \textbf{Rate} \\
\midrule
\multicolumn{4}{l}{\textit{High Actionability Models (Act. $>85\%$ in Table~\ref{tab:ncpquest})}} \\
Claude-Sonnet-4      & 157 & 32 & 11.5\% \\
GPT-4.1              & 157 & 35 & 16.6\% \\
Ablation-NoSpecialist& 157 & 34 & 16.6\% \\ 
\textbf{NutriOrion (Ours)}  & 132 & 23 & \textbf{12.1\%} \\
\midrule
\multicolumn{4}{l}{\textit{Moderate Actionability Baselines}} \\
RoundTable           & 114 & 14 &  8.8\% \\
React-Qwen-2.5-72B  & 126 &  6 &  4.8\% \\
Ensemble             & 157 &  5 &  3.2\% \\
SelfRefine-Qwen-2.5  & 155 & 11 &  7.1\% \\
\midrule
\multicolumn{4}{l}{\textit{\color{gray}Low Actionability Artifacts (Act. $<40\%$ in Table~\ref{tab:ncpquest})}} \\
\color{gray}Ablation-NoADIME      & \color{gray}157 & \color{gray} 0 & \color{gray} 0.0\% \\
\color{gray}React-LLaMa-3-70B    & \color{gray} 97 & \color{gray} 0 & \color{gray} 0.0\% \\
\color{gray}CoT-Qwen-2.5-72B      & \color{gray}157 & \color{gray} 1 & \color{gray} 0.6\% \\
\color{gray}CoT-LLaMA-3-70B       & \color{gray}157 & \color{gray} 4 & \color{gray} 2.5\% \\
\color{gray}SelfRefine-LLaMA-3    & \color{gray}155 & \color{gray} 0 & \color{gray} 0.0\% \\
\bottomrule
\end{tabular*}
\vspace{-1em}
\end{table}

\textbf{Results.} Table~\ref{tab:safety} reveals a systematic vulnerability we term the \textit{``Healthy Food Trap''}: high-performing models (NutriOrion, Claude, GPT) frequently recommend generally healthy foods (e.g., leafy greens, bananas) that are specifically contraindicated for pharmacotherapy subgroups, leading to violation rates of 11--17\%. This occurs because nutritional heuristics (``eat vegetables'') conflict with context-specific drug constraints (e.g., Vitamin K restriction for Warfarin).

While models like React-LLaMa-70B or Ablation-NoADIME appear ostensibly safer with 0.0\% violations, cross-referencing with Table~\ref{tab:ncpquest} exposes this as an artifact of low Actionability ($<40\%$). These models fail to prescribe concrete food items, outputting generic advice (e.g., ``eat healthy'') that bypasses safety filters by default. This \textit{passive safety} is clinically useless as it fails to guide patient behavior. NutriOrion (12.1\% violations) represents the only open-source architecture achieving active safety: it generates highly specific, actionable meal plans (Actionability 97.8\% in Table~\ref{tab:ncpquest}) while maintaining a safety profile comparable to frontier closed-source models (Claude-Sonnet-4 at 11.5\%). 

\textit{Failure Analysis.} 
Detailed breakdowns in Appendix Table~\ref{tab:safety-scenario} reveal that Warfarin interactions account for the majority of violations (21--57\% across high-actionability models). Despite NutriOrion's deterministic Safety Gate, the residual 12\% violation rate highlights a semantic gap in current pharmaceutical databases: explicit contraindications in Product Labels (e.g., ``avoid Vitamin K-rich foods'') do not always enumerate specific vegetables (spinach, kale), creating a terminology mismatch that rule-based filters cannot resolve. This underscores the urgent need for comprehensive drug-nutrient Knowledge Graphs (KG) to augment LLM reasoning.
%--------------------------------------------------------------------------------
% Table 4: Efficiency Comparison (Optional, for Discussion)
%--------------------------------------------------------------------------------
% \begin{table}[t]
% \centering
% \caption{Output Efficiency Comparison. NutriOrion achieves Excellent 
% documentation quality with significantly less output, demonstrating 
% efficiency in producing actionable recommendations.}
% \label{tab:efficiency}
% \begin{tabular}{@{}lccr@{}}
% \toprule
% \textbf{Model} & \textbf{Quality Tier} & \textbf{Avg. Output (chars)} & \textbf{Ratio} \\
% \midrule
% Claude-Sonnet-4    & Excellent & 17,925 & 6.7$\times$ \\
% GPT-4.1            & Excellent & 18,023 & 6.8$\times$ \\
% \textbf{NutriOrion}& Excellent &  2,663 & 1.0$\times$ \\
% React-Qwen-72B    & Excellent &  3,842 & 1.4$\times$ \\
% React-LLaMa-70B   & Good      &  2,105 & 0.8$\times$ \\
% \bottomrule
% \end{tabular}
% \vspace{-1em}
% \end{table}

\subsection{Clinical Appropriateness}
\label{sec:clinical_appr}

% NutriOrion Paper - Experiment Section: Evaluated by Registered Dietitians
To assess holistic clinical utility, the same two Registered Dietitians (RDs) from Section~\ref{sec:doc-quality} evaluated 30 randomly selected NutriOrion interventions using a structured survey instrument (Appendix~\ref{app:survey}). The survey aligns with NCP domains and post-stroke guidelines~\cite{stroke_nutrition}, comprising six dimensions: (A) Guideline Alignment, (B) Dietary Components, (C) Balance, (D) Clarity, (E) Safety (identification of unsafe foods), and (F) Overall Quality.

\textbf{Results.}
NutriOrion demonstrated strong clinical adherence, scoring highest in \textit{Guideline Alignment} (Mean=4.69/5) and \textit{Dietary Components} (4.66). \textit{Dietary Balance} (4.51) and \textit{Safety} (4.28) also achieved high ratings, confirming that the system effectively addresses multimorbidity requirements. Figure~\ref{fig: RD evaluation} in the appendix visualizes the evaluation profile. 

Notably, \textit{Overall Quality} scored 3.87. Open-text feedback clarified that while food recommendations were chemically accurate (Section~\ref{sec:personalization}) and specific (Section~\ref{sec:doc-quality}), the lack of "practical meal preparation guidance (e.g., recipes, culinary preparation methods)" lowered the adoption score. This distinguishes a logistical gap from the specificity gap solved by our architecture, highlighting that future clinical LLMs must extend beyond "what to eat" to "how to prepare and quantify it."

Despite this gap, quantitative analysis confirms clinical viability: 81.4\% of plans achieved clinical acceptability, and 64.4\% of RDs indicated willingness to sign off on them (Adoption Intention). Regarding safety, while 85.6\% of items received positive ratings, 37.3\% of evaluations noted missed subtle unsafe foods, corroborating the "Healthy Food Trap" findings in Section~\ref{sec:safety-audit}. 

Inter-annotator agreement varied substantially: highest for categorization tasks (79--97\%) but lowest for absolute \textit{safety red flag counts} (8\%). Qualitative audit reveals this stems from variance in annotation granularity rather than clinical dissent: one RD tended to flag individual contraindicated ingredients (e.g., "salt", "soy sauce" $\rightarrow$ count=2), while the other flagged the composite meal (e.g., "high sodium dish" $\rightarrow$ count=1). Crucially, evaluators achieved high consensus on the presence of safety risks, validating the overall safety scores despite the counting ambiguity.

% \begin{figure}
%     \centering
%     \includegraphics[width=0.8\linewidth]{NutriOrion26‘/figure/radar_chart.pdf}
%     \caption{RD Clinical Appropriateness Evaluation. Mean scores (1–5 scale) across six dimensions for NutriOrion outputs}
%     \label{fig:fig3}
% \end{figure}

\vspace{-0.5em}

\subsection{Responsiveness to Patient Profiles}
\label{sec:personalization}

To quantify personalization in generated nutrition interventions, we measured the alignment between patient health biomarkers and the nutritional composition of recommended foods. We mapped food items in generated interventions to Food and Nutrient Database for Dietary Studies (FNDDS) codes, the same coding system NHANES uses to obtain nutritional data, using the method from~\cite{yan2025dietai24}.  This enabled extraction of complete nutritional profiles for each recommended food. We selected five key health biomarkers: systolic blood pressure (BP), glucose, HbA1c, total cholesterol, and BMI. For each patient, we calculated Pearson correlation coefficients between health biomarkers and corresponding dietary nutrients: (1) systolic BP vs. sodium content, (2) glucose vs. sugar content, (3) HbA1c vs. sugar content, (4) total cholesterol vs. fat content, and (5) BMI vs. energy content. 

\begin{figure}[b]
    \centering
    \includegraphics[width=\linewidth]{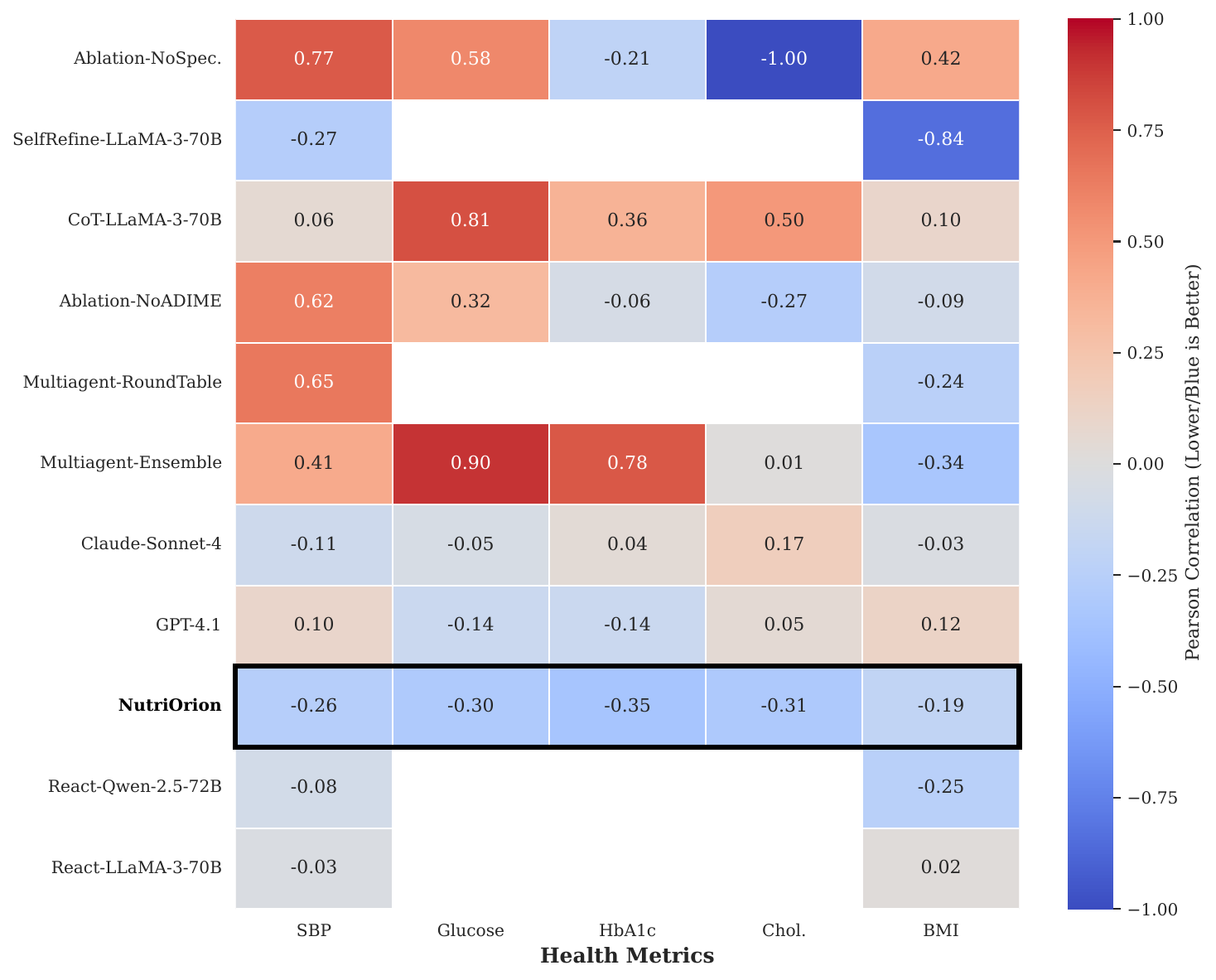}
    \caption{\textbf{Personalization Heatmap.} Pearson correlations between patient biomarkers and nutrient intake. Blue (Negative) indicates effective personalization (e.g., High BP $\rightarrow$ Low Sodium), while Red (Positive) implies unsafe recommendations. NutriOrion consistently achieves the strongest protective correlations across all five metrics, significantly outperforming baselines.}
    \label{fig:heatmap_personalization}
\end{figure}

The personalization hypothesis is that appropriate recommendations should exhibit negative correlations: when patients have elevated biomarkers (e.g., high BP), recommended foods should contain lower levels of risk nutrients (e.g., sodium). As visualized in Figure~\ref{fig:heatmap_personalization}, negative values (Blue) indicate high personalization quality, while positive values (Red) imply safety failures. NutriOrion demonstrates superior personalization by consistently achieving the strongest negative correlations across all five metrics (e.g., HbA1c: $-0.35$; Cholesterol: $-0.31$), appearing as the distinct deep blue region in the heatmap. In contrast, baseline models frequently exhibit unsafe positive correlations (red zones). Complete correlation results and comparative analysis are presented in Appendix~\ref{app:personalization}.

\subsection{Food Quality Assessment}
\label{sec:food_quality}

\begin{figure}
    \centering
    \includegraphics[width=\linewidth]{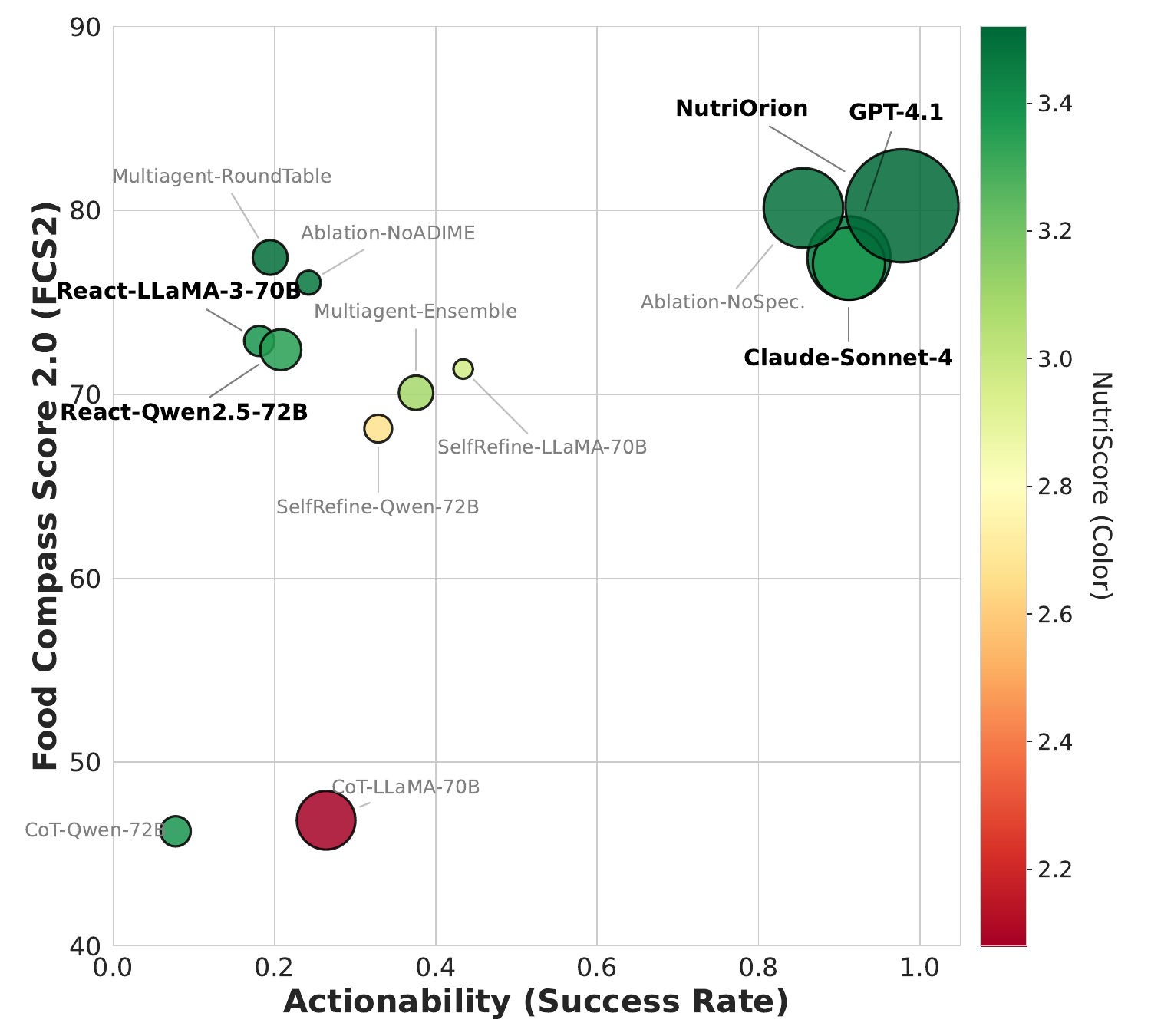}
\caption{\textbf{Multi-dimensional Performance (Bubble Chart).} 
    \textit{X-axis}: Actionability; \textit{Y-axis}: Food Compass Score 2.0 (Nutritional Quality); \textit{Bubble Size}: Dietary Diversity; \textit{Color}: NutriScore Health Gradient (Red$\rightarrow$Green).
    \textbf{NutriOrion} (top-right) achieves the optimal balance: high quality, actionable output, and high diversity. \textcolor{gray}{Gray labels} indicate unstructured baselines, as detailed in Table~\ref{tab:ncpquest}.
  }
    \label{fig:performance_bubble}
\end{figure}

To evaluate the nutritional quality of recommended foods, we employed Food Compass 2.0~\cite{barrett2024food}, a comprehensive food healthfulness ranking system that scores foods based on multiple nutritional attributes including nutrient composition, food processing level, and health outcomes associated with consumption. Food Compass 2.0 provides a rating score ranging from 0 to 100 for each FNDDS food code, with higher scores indicating greater nutritional quality. We leveraged the FNDDS food code mapping from Section~\ref{sec:personalization} to obtain Food Compass scores for each food item in the generated nutrition interventions. For each intervention, we calculated the average Food Compass score across all recommended food items, then aggregated these scores across all patient samples to obtain a mean nutritional quality score for each model. Figure~\ref{fig:performance_bubble} presents a multi-dimensional performance comparison, plotting actionability (x-axis) against Food Compass Score 2.0 (y-axis), with bubble size representing dietary diversity (unique food codes) and color indicating NutriScore-based health quality (green denotes higher scores).

\vspace{-0.5em}

Figure~\ref{fig:performance_bubble} visualizes the performance landscape, revealing distinct architectural behaviors. NutriOrion and GPT-4.1 occupy the Optimal Frontier (Top-Right), achieving both high quality (FCS $\approx$ 80) and high actionability ($>$ 0.85), proving that nutritional density can coexist with specific planning. In contrast, baselines like \textit{React-LLaMA} and \textit{RoundTable} fall into a "Conservative Trap" (Top-Left), achieving high quality scores ($\approx$75--78) but suffering from low Actionability ($<0.3$). This indicates a form of "Passive Quality", they achieve high scores by recommending generic healthy categories rather than specific meals, limiting their clinical utility. Meanwhile, CoT-based models mostly reside in the Failure Zone (Bottom-Left), failing in both dimensions with low-quality outputs (FCS $<50$, Red Bubble) and negligible actionability. The bubble size reveals that NutriOrion recommends a diverse range of foods, indicating varied recommendations tailored to individual patient contexts rather than repetitive suggestions. The combination of high food quality, high actionability, substantial diversity, and strong NutriScore demonstrates that NutriOrion successfully balances multiple competing objectives in nutrition intervention generation. NutriOrion's recommendations demonstrate clinically meaningful dietary improvements: reducing energy by 8\%, sodium by 9\%, and sugars by 12\%, while simultaneously increasing beneficial nutrients with potassium up 27\% and fiber up 167\%. These targeted modifications align with dietary guidelines for chronic disease management. Comprehensive nutritional composition analysis is presented in Appendix~\ref{app:nutrition_change}.

\section{Conclusion}
\vspace{-0.5em}

Our evaluation of NutriOrion on 330 multimorbid stroke patients demonstrates that specialized multi-agent architectures fundamentally outperform monolithic reasoning across safety, personalization, and clinical appropriateness dimensions, with controlled ablations revealing that both role specialization and structured ADIME workflows are necessary, as removing either degrades performance to baseline levels. The framework achieves 97.8\% actionability in generating specific food recommendations while maintaining 12.1\% drug-food violation rates comparable to frontier commercial APIs, with registered dietitians rating 81.4\% of plans as clinically acceptable and demonstrating robust personalization through consistent negative correlations (ranging from -0.26 to -0.35) between patient biomarkers and corresponding dietary risk nutrients across all evaluated health metrics. Building upon these foundational results, future work will extend the system in two important directions: incorporating precise portion size specifications by integrating standardized serving databases to provide quantitative nutrient intake control, and conducting prospective clinical trials to evaluate patient-reported outcomes including adherence rates, satisfaction scores, and longitudinal health improvements in real-world deployment settings. More broadly, NutriOrion enables on-demand dietary support accessible anywhere, increasing the reach of personalized nutrition intervention and allowing populations worldwide to benefit from advanced LLM capabilities and digital health tools for managing chronic diseases.

\newpage

\clearpage

\bibliographystyle{ACM-Reference-Format}
\bibliography{NutriOrion26/_reference}
\appendix

\clearpage

\appendix

\section{FHIR Interoperability Schema}
\label{sec:fhir_mapping}

To demonstrate the clinical deployment readiness of NutriOrion, we provide the explicit mapping schema between our system's internal ADIME-structured JSON output and the HL7 FHIR R4 \texttt{NutritionOrder} resource standard. This mapping ensures that the AI-generated interventions can be directly ingested by Electronic Health Record (EHR) systems without loss of semantic fidelity.

\begin{table}[h]
\caption{Full Mapping Schema: From NutriOrion ADIME Output to FHIR R4 NutritionOrder}
\label{tab:fhir_mapping_full}
\centering
\small % 保持小字体
\renewcommand{\arraystretch}{1.2}

% 将表格总宽度强制设为 \linewidth
% X 列会自动换行并填充剩余空间
% >{\raggedright\arraybackslash} 用于让 X 列左对齐，避免单词间距过大
\begin{tabularx}{\linewidth}{
    >{\raggedright\arraybackslash}X  % 第一列自动调整
    p{0.15\linewidth}                % 第二列保持固定
    >{\raggedright\arraybackslash}X  % 第三列自动调整
}
\toprule
\textbf{NutriOrion Internal Field} & \textbf{ADIME Stage} & \textbf{FHIR R4 Target Element} \\
\midrule
% 如果某些 textt 太长不换行，可以在关键点手动加 \allowbreak 或 \-
\texttt{hyper\-tension\_diet} (e.g., DASH) & Assessment & \texttt{oralDiet.type.coding} (e.g., SNOMED: 182922004) \\
\texttt{diabetes\_diet} (e.g., CarbCount) & Assessment & \texttt{oralDiet.type.coding} \\
\texttt{nutrient\_flags.sodium} & Diagnosis & \texttt{oralDiet.nutrient.modifier} (e.g., LOINC) \\
\texttt{nutrient\_flags.limit\_val} & Diagnosis & \texttt{oralDiet.nutrient.amount} \\
\texttt{inter\-vention.replace} & Intervention & \texttt{oralDiet.exclude\-FoodModifier} \\
\texttt{intervention.add} & Intervention & \texttt{oralDiet.food\-PreferenceModifier} \\
\texttt{medication.supplement} & Intervention & \texttt{supplement.type} \\
\texttt{health\_priority.rank} & Diagnosis & \texttt{priority} (Routine/Urgent) \\
\texttt{monitoring.frequency} & Monitoring & \texttt{oralDiet.schedule} \\
\bottomrule
\end{tabularx}
\end{table}

The mapping function $\Phi$ is implemented as a deterministic parser that validates the agent's output against the FHIR structure definition before final delivery.

\section{Implementation Details of Baselines}
\label{sec:appendix_baselines}

Experiments were conducted using vLLM for open-source models and official APIs for closed-source models. All agents operate at temperature \textbf{0.0} to ensure reproducibility in clinical reasoning.

\subsection{Single-Agent Configurations}
The single agent acts as a "Comprehensive Specialist".
\begin{itemize}
    \item \textbf{ReAct:} Standard loop with \texttt{max\_iter=50}. We implemented a \texttt{NoOpTool} to prevent early termination failures common in open-source models.
    \item \textbf{CoT:} Enforces a strict linear sequence: (1) Body Metrics $\rightarrow$ (2) Guidelines $\rightarrow$ (3) Meds $\rightarrow$ (4) Diet $\rightarrow$ (5) Report. No tool looping allowed.
    \item \textbf{Self-Refine:} A 3-stage chain sharing the LLM backbone: (1) \textit{Draft} (tool-enabled), (2) \textit{Critique} (text-only, $<200$ words to prevent context overflow), and (3) \textit{Refine} (tool-enabled).
\end{itemize}

\subsection{Multi-Agent Baseline Configurations}
All multi-agent baselines use GPT-4o-mini as the backbone.
\begin{itemize}
    \item \textbf{Ablations:}
    \textit{w/o ADIME} uses a hierarchical manager with the prompt: "There is no predefined workflow. Coordinate as you see fit."
    \textit{w/o Roles} homogenizes all agent prompts to a generic "Clinical Nutritionist" while keeping the DAG.
    
    \item \textbf{Round Table (AutoGen-style):} Consists of Physician, Dietitian, and Pharmacist with \texttt{allow\_delegation=True} to simulate free-form MDT meetings. A strict timeout is enforced to prevent infinite loops.
    
    \item \textbf{Ensemble (MedAgents-style):} Three parallel agents with distinct personas (\textit{Conservative, Proactive, Evidence-Based}) analyze independently. An Aggregator Agent merges outputs using majority voting for recommendations and union logic for safety warnings.
\end{itemize}

\section{Registered Dietitian Evaluation Survey}
\label{app:survey}

\begin{figure}
    \centering
    \includegraphics[width=0.8\linewidth]{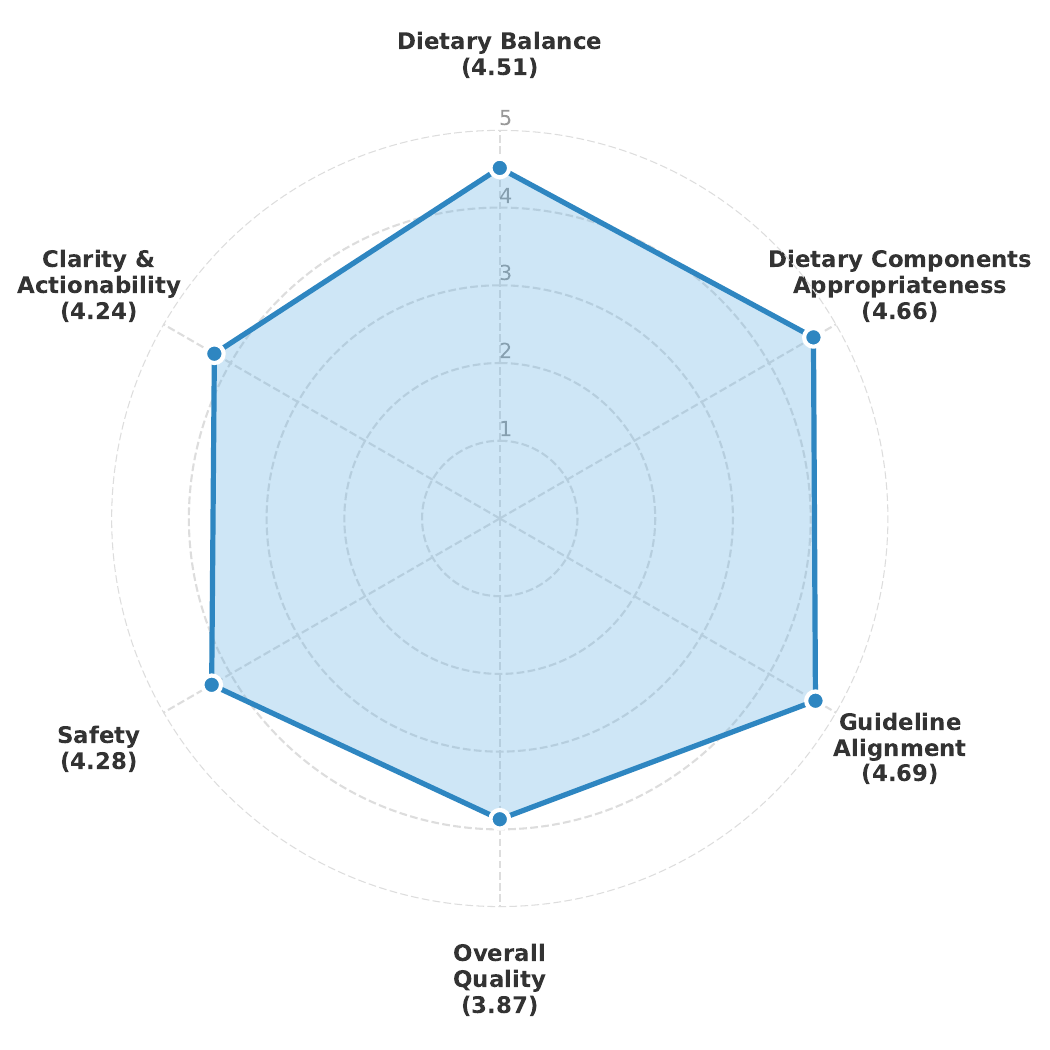}
    \caption{RD Clinical Appropriateness Evaluation. Mean scores (1–5 scale) across six dimensions for NutriOrion outputs}
    \label{fig: RD evaluation}
\end{figure}

Fig. \ref{fig: RD evaluation}RD Clinical Appropriateness Evaluation. Mean
scores (1–5 scale) across six dimensions for NutriOrion out-
puts
Below is the full content of the survey used for the dietitian review.

\noindent\rule{\linewidth}{2pt} % 顶部分隔线

\subsection*{Evaluation Survey}

\subsubsection*{Instructions}
Please evaluate the given nutrition plan for \textbf{this specific patient} using your professional judgment. Use a \textbf{5-point Likert scale} unless otherwise specified: \\
\textbf{1} = Strongly Disagree, \textbf{2} = Disagree, \textbf{3} = Neutral, \textbf{4} = Agree, \textbf{5} = Strongly Agree. \\
\textit{Provide N/A if a statement is not applicable, or insufficient information is available.}

% --- Section A ---
\subsection*{Section A: Patient Profile \& Guideline Alignment (GA)}
\begin{itemize}
    \item[\textbf{GA1.}] The plan is \textbf{consistent with the patient's health profile}, including diagnoses, anthropometrics, medications, allergies/intolerances. (1--5 / N/A)
    \item[\textbf{GA2.}] Overall, the plan \textbf{aligns with applicable evidence-based guidelines} for post-stroke care (e.g., ADA, DASH). (1--5)
\end{itemize}

% --- Section B ---
\subsection*{Section B: Appropriateness for Key Dietary Components (AP)}
\textit{Stem:} ``For the following dietary components, the AI's \textbf{restriction/intake recommendation is appropriate} given the patient's condition and guidelines.'' Rate each item:
\begin{itemize}
    \item[\textbf{AP1.}] Sodium (1--5 / N/A)
    \item[\textbf{AP2.}] Total Fat (\textit{considering saturated/trans-fat and cholesterol as part of total fat}) (1--5 / N/A)
    \item[\textbf{AP3.}] Total Energy (calories) (1--5 / N/A)
    \item[\textbf{AP4.}] Added Sugars (1--5 / N/A)
    \item[\textbf{AP5.}] Dietary Fiber (1--5 / N/A)
\end{itemize}

% --- Section C ---
\subsection*{Section C: Dietary Balance (DB)}
\begin{itemize}
    \item[\textbf{DB1.}] The overall dietary pattern is balanced across food groups (e.g., fruits/vegetables, whole grains, low-fat dairy, lean proteins, healthy fats). (1--5)
\end{itemize}

% --- Section D ---
\subsection*{Section D: Clarity \& Actionability (CL)}
\begin{itemize}
    \item[\textbf{CL1.}] The patient-facing summary is clear, concise, and understandable. (1--5)
    \item[\textbf{CL2.}] The ``Why'' explanations are accurate and evidence-informed. (1--5)
    \item[\textbf{CL3.}] The diet orders/instructions (limits/substitutions/additions) are specific and actionable (e.g., portions, frequency, concrete examples). (1--5)
\end{itemize}

% --- Section E ---
\subsection*{Section E: Safety (SF)}
\textit{Answer Yes/No. If ``No,'' a brief comment is required.}

\begin{description}
    \item[\textbf{Safety Red Flags (Definition):}] Any recommendation that conflicts with the patient's diagnoses, medications, or allergies/intolerances; any intake that exceeds/under-shoots guideline thresholds posing clear risk; or any explicit drug--nutrient--disease incompatibility.
\end{description}

\begin{itemize}
    \item[\textbf{SF1.}] \textbf{Identification:} The plan correctly identifies unsafe foods in the patient's existing diet. (Yes/No) \\
    \textit{If No, please explain:} \underline{\hspace{0.6\linewidth}}
    
    \item[\textbf{SF2.}] \textbf{Substitutions:} Proposed substitute foods are functionally comparable (meet the same dietary intent) and safe. (Yes/No) \\
    \textit{If No, please explain:} \underline{\hspace{0.6\linewidth}}
    
    \item[\textbf{SF3.}] \textbf{Additions:} Newly recommended foods are safe and effective for the patient. (Yes/No) \\
    \textit{If No, please explain:} \underline{\hspace{0.6\linewidth}}
    
    \item[\textbf{SF4.}] \textbf{Contraindications:} The plan does not introduce drug--nutrient or disease-related contraindications. (Yes/No) \\
    \textit{If No, please explain:} \underline{\hspace{0.6\linewidth}}
    
    \item[\textbf{SF5.}] Number of Safety Red Flags identified: \underline{\hspace{1cm}} (integer; 0 if none)
\end{itemize}

% --- Section F ---
\subsection*{Section F: Overall Quality (OQ)}
\begin{itemize}
    \item[\textbf{OQ1.}] Overall, this AI-generated plan is clinically acceptable for this patient. (1--5)
    \item[\textbf{OQ2.}] (Adoption intention) I would be comfortable signing/co-signing this plan. (1--5)
    \item[\textbf{OQ3.}] Other comments or risks not covered above: 
\end{itemize}

% --- Section G & H ---
\subsection*{Section G: Workload (WL)}
\begin{itemize}
    \item[\textbf{WL1.}] Time spent reviewing this plan (minutes): \underline{\hspace{1.5cm}}
\end{itemize}

\subsection*{Section H: Open Comments}

\noindent\rule{\linewidth}{2pt} % 底部分隔线

\section{NCP-QUEST Evaluation Criteria}
\label{app:ncpquest}

In our expert evaluation (Section~\ref{sec:doc-quality}), Registered Dietitians assessed the generated nutrition documentation based on the following four dimensions derived from the NCP-QUEST framework~\cite{lewis2022ncpquest}:

\begin{itemize}[leftmargin=*]
    \item \textbf{Nutrition Diagnosis (D):} Evaluates the accuracy in identifying patient-specific nutritional problems. A score of 2 requires the explicit prioritization of the most urgent nutritional issue based on the provided clinical data.
    \item \textbf{Nutrition Intervention (I):} Assesses the clinical appropriateness of the structured actions. Recommendations must use standard terminology (e.g., ``continue,'' ``replace,'' ``add'') to receive a full score.
    \item \textbf{D$\rightarrow$I Link:} Measures the logical consistency between the diagnosis and the prescribed intervention. The intervention must directly address the etiology or signs/symptoms identified in the diagnosis.
    \item \textbf{Specificity:} Evaluates the level of actionable detail. A score of 2 is awarded only if recommendations prescribe specific food items (e.g., ``grilled salmon,'' ``spinach'') rather than generic categories (e.g., ``healthy protein,'' ``vegetables'').
\end{itemize}

Each dimension is scored on a discrete scale:
\begin{itemize}[nosep]
    \item \textbf{0 (Not Documented):} The criteria are missing or incorrect.
    \item \textbf{1 (Partially Documented):} The criteria are present but lack specific detail or full clinical alignment.
    \item \textbf{2 (Adequately Documented):} The criteria are fully met with specific, actionable, and clinically accurate details.
\end{itemize}

%-------

%--------------------------------------------------------------------------------
% Table 3: Safety Violations by Scenario (For Appendix)
%--------------------------------------------------------------------------------
\begin{table*}[t]
\centering
\caption{Drug-Food Interaction Violations by Clinical Scenario (DailyMed-based). 
The three scenarios represent high-risk drug-food interactions prevalent in 
stroke patients. Warfarin users face the highest violation rates across all 
high-quality models due to the ``healthy food trap'' where vitamin K-rich 
vegetables (spinach, broccoli, kale) are commonly recommended.}
\label{tab:safety-scenario}
\begin{tabular}{@{}l|ccc|ccc|ccc@{}}
\toprule
& \multicolumn{3}{c|}{\textbf{Warfarin (n=28)}} 
& \multicolumn{3}{c|}{\textbf{K-Sparing Diuretics (n=15)}} 
& \multicolumn{3}{c}{\textbf{Statins + Stroke (n=114)}} \\
\textbf{Model} & Eval & Viol & Rate & Eval & Viol & Rate & Eval & Viol & Rate \\
\midrule
Claude-Sonnet-4    & 28 &  6 & 21.4\% & 15 & 11 & 73.3\% & 114 & 1 & 0.9\% \\
\textbf{NutriOrion}& 26 &  9 & 34.6\% & 12 &  7 & 58.3\% &  94 & 0 & 0.0\% \\
GPT-4.1            & 28 & 13 & 46.4\% & 15 & 11 & 73.3\% & 114 & 2 & 1.8\% \\
Ablation-NoSpec.   & 28 & 16 & 57.1\% & 15 & 10 & 66.7\% & 114 & 0 & 0.0\% \\
Single-Qwen-72B    & 23 &  5 & 21.7\% & 10 &  1 & 10.0\% &  93 & 0 & 0.0\% \\
SelfRefine-Qwen    & 28 &  7 & 25.0\% & 15 &  3 & 20.0\% & 112 & 1 & 0.9\% \\
RoundTable         & 21 &  6 & 28.6\% & 12 &  4 & 33.3\% &  81 & 0 & 0.0\% \\
\bottomrule
\end{tabular}
\vspace{0.5em}
\\
\small\textit{Note: The potassium-sparing diuretic scenario shows uniformly 
high violation rates (20--73\%) because bananas, avocados, and tomatoes---common 
``heart-healthy'' recommendations---are contraindicated for hyperkalemia risk.}
\end{table*}

\section{Personalization Correlation Analysis}
\label{app:personalization}

Figure~\ref{fig:heatmap_personalization} presents a heatmap visualization of Pearson correlations between patient health biomarkers (systolic BP, glucose, HbA1c, total cholesterol, BMI) and the corresponding dietary nutrients in recommended foods (sodium, sugar, sugar, fat, calories respectively) across different models. The color scale ranges from dark blue (strong negative correlation) to dark red (positive correlation), where more negative values indicate better personalization, as a well-personalized system should reduce risk nutrients when patients have elevated health biomarkers. NutriOrion (highlighted) consistently achieves the strongest negative correlations across four of five health metrics compared to all baselines, with correlation coefficients ranging from -0.26 to -0.35 for systolic BP, glucose, HbA1c, and total cholesterol, indicating robust personalization where the multi-agent framework successfully interprets these health biomarkers and adapts dietary recommendations accordingly. Notably, several baseline models exhibit positive correlations for certain metrics (e.g., CoT-LLaMA-3-70B shows 0.81 for glucose, Multiagent-Ensemble shows 0.78 for HbA1c), which represents unsafe recommendations that could worsen health conditions by suggesting foods high in risk nutrients for patients with elevated biomarkers. For BMI, NutriOrion shows -0.19 correlation, which is not the strongest among compared methods, as SelfRefine-LLaMA-3-70B (-0.84) and Multiagent-Ensemble (-0.34) demonstrate stronger negative correlations; however, as discussed in Section~\ref{sec:doc-quality}, these models produced substantially lower actionability rates and generated fewer specific food recommendations, resulting in limited samples for correlation calculation, making their BMI correlations potentially unrepresentative of consistent personalization capability. Overall, NutriOrion's balanced performance across all five health metrics demonstrates its ability to provide comprehensive, safe personalization for multimorbid patients.

\section{Nutritional Composition Changes}
\label{app:nutrition_change}

\begin{figure}[b]
    \centering
    \includegraphics[width=\linewidth]{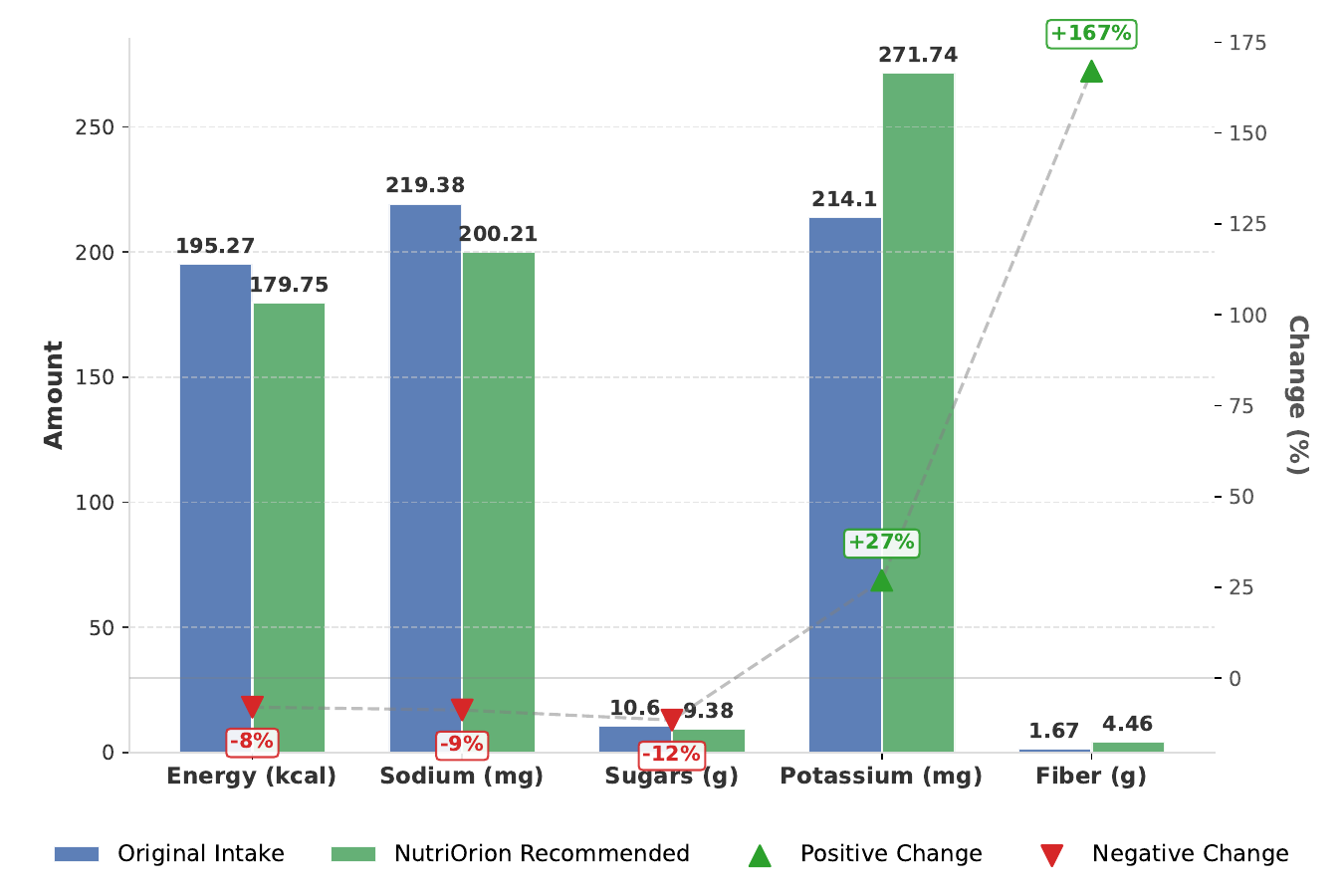}
    \caption{Average dietary nutrients from original foods and NutriOrion recommendations. Bars show absolute amounts, and markers indicate percent change relative to original intake.}
    \label{fig:labeled_nutrient_chart}
\end{figure}

Figure~\ref{fig:labeled_nutrient_chart} compares the average dietary nutrient content between patients' original food intake (blue bars) and NutriOrion's recommended foods (green bars) across five key nutrients, with bars showing absolute amounts and markers indicating percent change relative to original intake (green upward triangles for positive changes, red downward triangles for negative changes). NutriOrion's recommendations demonstrate clinically meaningful dietary modifications aligned with chronic disease management guidelines by reducing intake of risk-associated nutrients: energy decreases by 8\% (from 195.27 to 179.75 kcal), sodium decreases by 9\% (from 219.38 to 200.21 mg), and sugars decrease by 12\% (from 10.6 to 9.38 g), which are critical reductions for managing obesity, hypertension, and diabetes respectively. Simultaneously, NutriOrion substantially increases intake of beneficial nutrients, with potassium increasing by 27\% (from 214.1 to 271.74 mg) and dietary fiber increasing by 167\% (from 1.67 to 4.46 g); higher potassium intake benefits both hypertension and cardiovascular health by counteracting sodium's effects, while increased fiber improves glycemic control in diabetes and promotes satiety for weight management. These nutrient shifts demonstrate that NutriOrion does not simply restrict intake across all categories but rather makes targeted, evidence-based adjustments, reducing risk-associated nutrients while simultaneously increasing protective nutrients, reflecting sophisticated understanding of dietary guidelines for multimorbidity management. The substantial fiber increase (167\%) is particularly noteworthy, as fiber is often deficient in typical diets and plays multiple beneficial roles in chronic disease management, validating that NutriOrion generates actionable dietary interventions that translate high-level clinical objectives into concrete, beneficial nutritional changes.

% \begin{table}[t]
%   \centering
%   \caption{Pearson correlation between EHR health metrics and mean dietary change (Personalization). More negative values indicate better personalization.}
%   \label{tab:personalization}
  
%   \begin{tabular}{l|ccccc}
%     \toprule
%     \textbf{Method} & \textbf{SBP} & \textbf{Glucose} & \textbf{HbA1c} & \textbf{Chol.} & \textbf{BMI} \\
%     \midrule
%     Ablation-NoSpec.           & 0.77 & 0.58 & -0.21 & -1.00 & 0.42 \\
%     SelfRefine-LLaMA-3-70B     & -0.27 & N/A  & N/A   & N/A   & -0.84 \\
%     SelfRefine-Qwen2.5-72B     & N/A  & N/A  & N/A   & N/A   & N/A  \\
%     CoT-Qwen-2.5-72B           & N/A  & N/A  & N/A   & N/A   & N/A  \\
%     CoT-LLaMA-3-70B            & 0.06 & 0.81 & 0.36  & 0.50  & 0.10 \\
%     Ablation-NoADIME           & 0.62 & 0.32 & -0.06 & -0.27 & -0.09 \\
%     Multiagent-RoundTable      & 0.65 & N/A  & N/A   & N/A   & -0.24 \\
%     Multiagent-Ensemble        & 0.41 & 0.90 & 0.78  & 0.01  & -0.34 \\
%     Claude-Sonnet-4            & -0.11 & -0.05 & 0.04 & 0.17 & -0.03 \\
%     GPT-4.1                    & 0.10 & -0.14 & -0.14 & 0.05 & 0.12 \\
%     \midrule
%     \textbf{NutriOrion}        & \textbf{-0.26} & \textbf{-0.30} & \textbf{-0.35} & \textbf{-0.31} & \textbf{-0.19} \\
%     \midrule
%     React-Qwen-2.5-72B         & -0.08 & N/A  & N/A   & N/A   & -0.25 \\
%     React-LLaMA-3-70B          & -0.03 & N/A  & N/A   & N/A   & 0.02 \\
%     \bottomrule
%   \end{tabular}%
  
% \end{table}

% \bibliographystyle{ACM-Reference-Format}

% \appendix
% \input{iclr2026/section/7_appendix}

\end{document}